%% file: Hierarchies-V3.tex
\documentclass[11pt,english,twoside]{article}
\pdfoutput=1
\usepackage[none]{hyphenat}
\usepackage{jheppub}
\usepackage[numbers]{natbib}
\usepackage{lscape}
\usepackage{amsmath}
\usepackage{xcolor}
\usepackage{graphicx}
\usepackage{dcolumn}
\usepackage{bm}
\usepackage{tikz}
\usetikzlibrary{decorations.markings}

\def\-{\hphantom{-}}

\def\s2{\frac{1}{\sqrt2}}

\def\beq{\begin{equation}}
\def\eeq{\end{equation}}
\def\bea{\begin{align}}
\def\eea{\end{align}}
\def\beqa{\begin{eqnarray}}
\def\eeqa{\end{eqnarray}}

\def\mg{m_{3/2}}
\def\mg2{m^2_{3/2}}

\def\Dsl{\,\raise.15ex\hbox{/}\mkern-13.5mu D} 

\def\Re{\text{Re}}
\def\Im{\text{Im}}

\newcommand{\eq}[1]{\begin{equation}
                     \begin{split} #1 \end{split}
                     \end{equation}}

\begin{document}

\title{Leaving the Swampland: Non-geometric fluxes and the Distance Conjecture}

\author[a]{Nana Cabo Bizet,}
\author[b]{Cesar Damian,} 
\author[a]{Oscar Loaiza-Brito}
\author[a]{and Damian Mayorga Pe\~na}

\affiliation[a]{Departamento de F\'isica, Universidad de Guanajuato, \\
Loma del Bosque No. 103 Col. Lomas del Campestre C.P 37150 Le\'on, Guanajuato, Mexico.}
\affiliation[b]{Departamento de Ingenier\'ia Mec\'anica, Universidad de Guanajuato, \\
Carretera Salamanca-Valle de Santiago Km 3.5+1.8 Comunidad de Palo Blanco, Salamanca, Mexico}

\emailAdd{nana@fisica.ugto.mx}
\emailAdd{cesaredas@fisica.ugto.mx}
\emailAdd{oloaiza@fisica.ugto.mx}
\emailAdd{dmayorga@fisica.ugto.mx}

\date{\today}

\abstract{We study a Type IIB isotropic toroidal compactification with non-geometric fluxes. Under the assumption of a hierarchy on the moduli, an effective scalar potential is constructed showing a runaway direction on the real part of the K\"ahler modulus while the rest of 
the moduli are stabilized. For the effective model to be consistent it is required that  displacements in the field space are finite. Infinite distances in field space would imply a breakdown in the hierarchy assumption on the moduli.
In this context, the Swampland Distance Conjecture is satisfied  suggesting the possibility of leaving or entering the Swampland by a parametric control of the fluxes. This is achieved upon allowing the non-geometric fluxes to take fractional values. In the process we are able to compute the cut-off scale below which the theory is valid, completely depending on the flux configuration. We also report on the appearance of a discrete spectrum of values for the string coupling at the level of the effective theory.
}
\arxivnumber{}


\maketitle


\section{Introduction}

\input{intro}

\section{Swampland criteria in type IIB toroidal compactifications}

\input{dS}

\section{Inclusion of non-geometric fluxes and the Moduli Hierarchy Assumption}

\input{modulih}

\section{Hierarchy and the Swampland Distance Conjecture}
\input{hierarchys}

\subsection{A toy example: Quintessence and the swampland}

\input{quintessence}

\section{Final comments}

\input{conclusions2}

\begin{center}
{\bf Acknowledgments}
\end{center}
We thank Anamar\'{\i}a Font, Xin Gao, Fernando Quevedo, Pramod Shukla and Ivonne Zavala for interesting and useful discussions on related topics. C.D. is supported by CONACyT grant number 66210, N.C.B. thanks the support of DAIP-UG 2019 
and CONACyT Project A1-S-37752, O. L.-B. and D. M. P. are supported by CONACyT grant number 258982 CB-2015-01 and by DAIP-UG 2019.

\appendix

\section{Hierarchies}
\label{sec:hierarchies}
\input{appendix1}

\section{Fixing the notation}
\label{sec:notation}
\input{appendix2}

\bibliographystyle{JHEP}
\bibliography{References}

\end{document}

%% file: intro.tex
The Swampland program \cite{Vafa:2005ui, Ooguri:2006in,  Obied:2018sgi, Agrawal:2018own,Garg:2018reu, Ooguri:2018wrx} has received a great 
deal of attention in the last few years (for reviews see e.g. \cite{Brennan:2017rbf, Palti:2019pca}). The proposed criteria to discern wether or not an effective theory is compatible with a quantum gravity theory leads us to the possibility to understand important issues concerning the construction of effective theories directly from string theory or inspired by it \cite{Yamazaki:2019ahj}. The possible microscopic origin of the criteria is also an opportunity to question wether our knowledge about dimensional reduction and compactification in generic scenarios is complete \cite{Marsh:2018kub, Danielsson:2018qpa}.\\

One issue has to do with the the validity of the Swampland criteria as a way to have a well-defined boundary in field theory, separating those compatible effective theories with string theory to those which are not. If true, one should be able to  cross it in both directions \cite{Murayama:2018lie} (see \cite{Andriot:2019wrs, Dvali:2018jhn, Garg:2018zdg, Hebecker:2018vxz, Wang:2018kly, Brahma2019, Ashoorioon:2018sqb, Motaharfar:2018zyb, Hamaguchi:2018vtv, Das:2018hqy, Kallosh:2019axr, Raveri:2018ddi, Acharya:2018deu, Achucarro:2018vey, Akrami:2018ylq, Kachru:2018aqn, Conlon:2018eyr,Brahma:2019mdd,Brahma:2019kch} for many different tests). Entering the Swampland seems to be easy by departing from an effective theory constructed from string theory  restricted to many of the self-consistent aspects of string theory at high energies
and by adding many extra assumptions. \\

One question arises however, about the meaning of leaving the Swampland. Let us say that one has an effective theory constructed by implementing some set of assumptions, inspired by a string construction. The number of assumptions could lead to a model violating some of the Swampland criteria. This is reflected by the apparent possibility to extend the model to transplanckian scales or to scales below the validity of the field theory. In this case the model is not just incomplete from the quantum gravity perspective but inconsistent at the level of field theory. However, suppose it is possible to identify some set of assumptions whose removal allows to fulfill the Swampland criteria and enter the Landscape.  Is the final model compatible with string theory? Are the removed assumptions a way to trace back consistent models? If all the above is true, one can establish a way to identify those assumptions that can be relaxed in an effective theory by entering the Landscape. In the process one would learn more about the UV completions of the effective theory.\\

Of particular interest becomes the construction of effective theories inspired by string theory, or as coined in \cite{Palti:2019pca}, string-inspired models. On those, the direct construction of a model from a concrete string theory is not completely known. In consequence the set of assumptions is an arbitrary election in their construction, playing in some cases, an important role in the consistency of the theory. \\

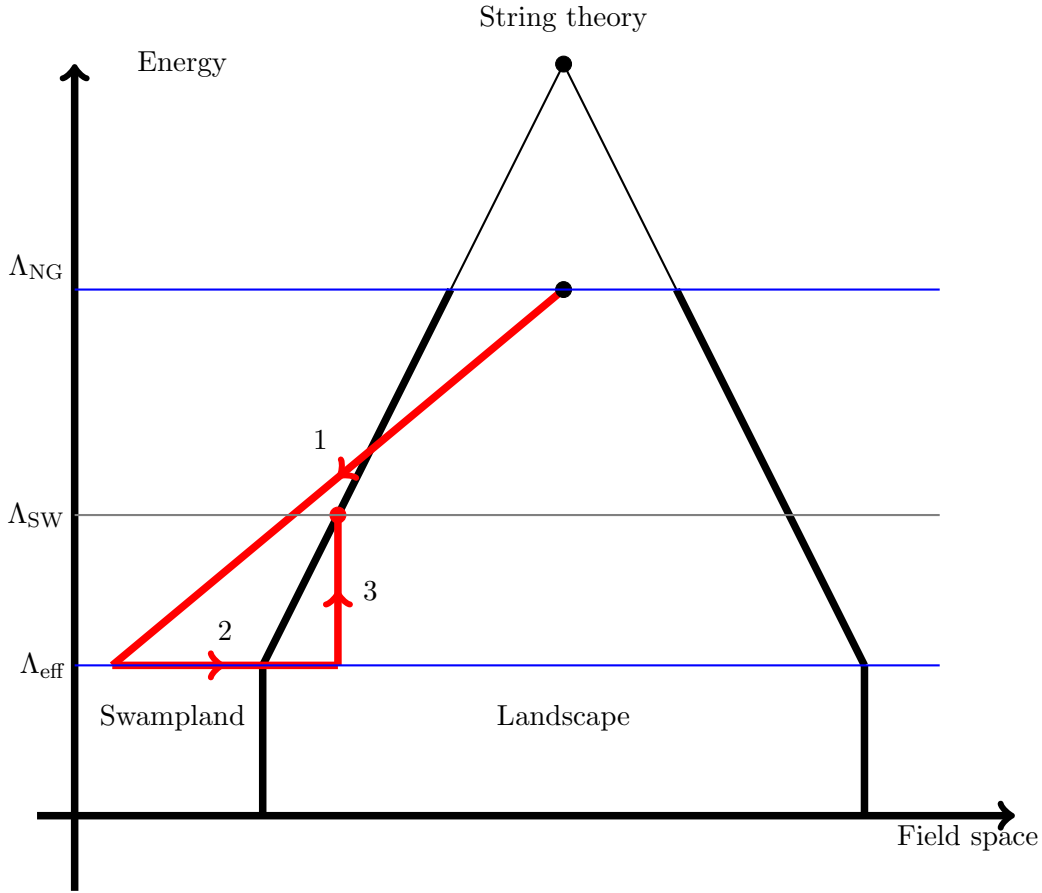
\begin{figure}
\centering
\begin{tikzpicture}
\draw[black, thick] (6,8) -- (2,0);
\draw[black, thick] (6,8) -- (10,0);
\draw[black,line width=1mm] (10,0) -- (7.5,5);
\draw[black,line width=1mm] (10,0) -- (10,-2);
\draw[black,line width=1mm] (4.5,5) -- (2,0);
\draw[black,line width=1mm] (2,0) -- (2,-2);
\begin{scope}[very thick,decoration={
    markings,
    mark=at position 0.5 with {\arrow{>}}}
    ] 
\draw[red, line width =1mm, postaction={decorate}] (0,0) -- (3,0);
\draw[red, line width =1mm, postaction={decorate}] (6,5) -- (0,0);
\draw[red, line width =1mm, postaction={decorate}] (3,0) -- (3,2);

\end{scope}
\filldraw[black] (6,8) circle (3pt);
\filldraw[black] (6,5) circle (3pt);
\filldraw[red] (3,2) circle (3pt);
\draw[black, thick,->,line width=1mm] (-0.5,-3) -- (-0.5,8);
\draw[black, thick,->,line width=1mm] (-1,-2) -- (12,-2);
\draw[gray, thick] (-0.5,2) -- (11,2);
\draw[blue, thick] (-0.5,0) -- (11,0);
\draw[blue, thick] (-0.5,5) -- (11,5);
\node[right] at (0.2,8) {Energy};
\node[left] at (-0.5,5.3) {$\Lambda_{\text{NG}}$};
\node[left] at (-0.5,2.0) {$\Lambda_{\text{SW}}$};
\node[left] at (-0.5,0.0) {$\Lambda_{\text{eff}}$};
\node[left] at (3,3) {1};
\node[above] at (1.5,0.2) {2};
\node[right] at (3.2,1.0) {3};
\node[above] at (6,8.3) {String theory};
\node[above] at (6,-1) {Landscape};
\node[above] at (0.8,-1) {Swampland};
\node[right] at (10.3,-2.3) {Field space};
\end{tikzpicture}
 \caption{Based in the diagram introduced in \cite{Palti:2019pca} we show the status of the model under consideration. By taking a string-inspired model resulting from a non-geometric flux compactification on an isotropic torus and by introducing strong constraints  at the scale $\Lambda_{NG}$ (represented by the arrowed trajectory 1 in the field space) we can end up with an effective theory in the Swampland in which the energy scale factor is unlimited (typically according to the value of some model-dependent parameters) below and above  the effective field theory scale  $\Lambda_{\rm eff}$. The more separated the trajectory from the line center, the more number of taken assumptions. By removing some set of assumptions (in principle different from those taken in trajectory 1) one can re-enter into the Landscape (trajectory 2), making possible to construct an effective theory valid till some scale $\Lambda_{SW}>\Lambda_{\rm eff}$ (trajectory 3).}
 \label{figure1}
\end{figure}

The presence of many assumptions could lead us to an effective theory violating some consistency checks such as the Swampland criteria, pushing the model directly into the Swampland. As proposed, if one of the Swampland criteria is violated the effective model cannot be completed in the UV, pointing out the presence of a model incompatible with some extension to quantum gravity, or as in the case, to string theory.  A second case could just lead us to a model valid untill some scale $\Lambda_{SW}$ above which some corrections or removal of some taken assumptions need to be implemented in order to have a model compatible with string theory. See Figure \ref{figure1}.\\

Consider the case of a  scalar potential with a runaway direction. In this case the refined dS conjecture is fulfilled over a finite range for the modulus field, such that infinite trajectories are limited by the distance conjecture. The relationship between these two Swampland criteria has been intensively studied in recent times \cite{Brown:2015iha, Corvilain:2018lgw, Scalisi:2018eaz, Junghans:2018gdb, Agrawal:2018rcg, Olguin-Tejo:2018pfq} indicating a link between them in terms of modular symmetries \cite{Gonzalo:2018guu} and the presence of non-perturbative objects such as instantons \cite{Marchesano:2019luw, Font:2019cxq}. Those works establish important advances in the search for the microscopic origin of the Swampland criteria (see also \cite{Lee:2019xtm, vandeBruck:2019vzd, Cole:2018emh}).\\

In this paper we explore the relation between the existence of a finite distance in field space and specific flux configurations by studying a simple string-inspired model consisting of a toroidal compactification of type IIB in the presence of non-geometric fluxes \cite{Shelton:2005cf, Shelton:2006fd, Wecht:2007wu}. These fluxes have been considered in the construction of effective models in order to generate a superpotential depending on all moduli fields, including K\"ahler moduli (see \cite{Plauschinn:2018wbo} for a review) by assuming the presence of T-duality at the level of the effective theory. Despite some significant success, mainly in the search for stable vacua with stabilized moduli \cite{Danielsson:2012by, Blaback:2013ht, Damian:2013dq, Damian:2013dwa, Blumenhagen:2015kja, Blumenhagen:2015xpa}, non-geometric fluxes lack for a complete global construction from the ten-dimensional perspective \cite{Bena:2017uuz, Blumenhagen:2015kja} (see however \cite{Blumenhagen:2015lta} for a proposal based on double field theory). Their incorporation into compactification models usually requires of the imposition of some set of plausible constraints, making these type of constructions perfect examples of the so called string-inspired models. The most usual assumptions involve the extension of tadpole and Bianchi identities to include
the corresponding T-dual fluxes, quantization of non-geometric fluxes and a null back-reaction by non-geometric fluxes on the internal geometry implying the assumption of a well-defined internal volume.\\

Non-geometric fluxes have also been studied in the context of the flux-scaling scenario in order to have some parametrical control to generate almost flat directions in moduli space, testing wether inflationary directions and stabilization of all moduli come along \cite{Blumenhagen:2015kja, Blumenhagen:2015xpa, Font:2017tuu, Blumenhagen:2015jva}. This approach was followed in the context of F-term axion monodromy inflation \cite{Marchesano:2014mla, Blumenhagen:2014nba} (see \cite{Blumenhagen:2018hsh, Blumenhagen:2018nts, Blumenhagen:2017cxt, Bizet:2016paj}  for relations among  the flux-scaling scenario, hierarchy on moduli mass and  the Swampland). So far all results seem to enforce an intriguing idea:  for inflationary directions to be present and to have a parametrical control over the different scales, fractional fluxes are required. In the context of the Swampland criteria, the above could be interpreted as a way to enter the Swampland, i.e. by adding an extra assumption concerning the presence of non-integer fluxes.\\

Within this context, we study a scenario of non-geometric flux compactification on an isotropic $T^6$  with orientifold three-planes \cite{Aldazabal:2006up} as an string-inspired effective model. Besides the inherent assumption about the validity of T-duality in four-dimensions, we also assume a hierarchy on the masses for the complex structure and the axio-dilaton against the K\"ahler modulus mass. We concentrate on a particular solution for which the superpotential component (depending on the K\"ahler moduli and the vevs for the complex structure and the axio-dilaton) vanishes \cite{Landete:2018kqf}. By this considerable increment on the number of assumptions, we find an analytical solution in which the scalar potential exhibits a runaway direction on the real component of the K\"ahler modulus $(\tau)$ resembling some characteristics of the KKLT-scenario\footnote{See \cite{Hamada:2019ack, Gautason:2019jwq, Blumenhagen:2019qcg, Hamada:2018qef,CaboBizet:2016qsa} for recent studies on KKLT scenario and the Swampland criteria.} before the inclusion of anti-branes.  There are some important results in our model we want to stress out here:\\
\begin{enumerate}
\item Compatibility with the Hierarchy Assumption on moduli fields forces the existence of a range in the field space for $\tau$ as suggested in \cite{Landete:2018kqf}. Since such a hierarchy comes from an appropriate selection of fluxes we conclude that in this case, the constraints on the flux configuration can be interpreted as the microscopic origin of the distance conjecture constraint. Notice as well that this provides the model with an essential feature since the volume is restricted to a range as expected to the geometric back-reaction of non-geometric fluxes.
\item The above condition defines a scale of energy $\Lambda_{SW}$ up to which the model is valid and it depends purely on non-geometric fluxes.
\item Due to the number of constraints, the effective scalar potential depends only on 2 non-geometric fluxes.
\item For different flux configurations (eight thousand) we find numerical evidence that the string coupling $s_0$ always acquires discrete values (see Figure \ref{fig:P3p}).
\item Taking integer values for all fluxes leads us to an incompatible effective theory with $\Lambda_{SW}>M_{s}$ and with an internal volume smaller than $1/M^6_s$, with $M_s$  the string scale.
\item Only by considering fractional values for non-geometric fluxes, the model is consistent and the distance conjecture is satisfied for a scale $\Lambda_{SW}$ below $M_{s}$. Similarly, the scale's hierarchies $M_s>M_{KK}>M_{U,S}$ are also accomplished and more importantly, it is possible to have a parametrical control by fluxes. We comment about fractional fluxes in our conclusions.
\end{enumerate}
By all the above we show a specific procedure which takes an effective model out of the Swampland by removing some, in principle, essential assumptions. Wether this mechanism is an available and general option to construct consistent effective models within the Swampland project is something we want to discuss.\\

This work is organized as follows: In Section 2 we review some toy models on which we based our proposal, stressing out important characteristics of having a runaway direction to check for consistency with the Swampland conjectures, specifically the refined dS and the distance conjectures. In Section 3 we describe the consistency conditions of the model by assuming a hierarchy on the moduli fields. The main part of our work is described in Section 4, where we present an analytical solution for fixing vevs of some moduli fields. This establishes a way to construct a scalar potential with a runaway direction which allows us to leave the Swampland by relaxing the condition of having integer-valued fluxes.  We discuss the consistency of the model, the Swampland criteria and the implications in the flux scaling scenario. Additionally we present numerical evidence supporting our assertions and a simple example in which the field $\tau$ acquires a small range in the field space parametrically controlled by the flux configuration. Finally we present our conclusions. In Appendix A we write the conditions for the scalar potential extrema in terms of the superpotential covariant derivates and Appendix B is devoted to an exhaustive discussion of the notation used throughout the paper.

%% file: dS.tex
In this Section we review some of the main features  that effective scalar field theories possess when constructed directly from a ten-dimensional string theory. Effective models  with a runaway direction constructed by  a toroidal compactification of Type IIB string theory,  constitute interesting models for which the Swampland refined de Sitter and Distance criteria are satisfied. Those
are usually driven by the real part of K\"ahler modulus.\\

\subsection{Toy model 1: GKP and the Swampland criteria}

Let us start by considering the Gukov-Vafa-Witten (GVW) superpotential ${\cal W}$ derived from a 6-dimensional isotropic toroidal compactification of Type IIB string theory. The superpotential depends only on two moduli, the complex structure  $U=u+{\rm i}v$ and the complex axio-dilaton $S=s+{\rm i}c$,
\eq{
{\cal W}(U,S)= P_1(U)-{\rm i}SP_2(U),
}
with
\begin{align}
P_1(U)&=f^0+3{\rm i}f_\ast U -3f^\ast U^2-{\rm i}f_0 U^3, \label{eq:p1}\\
P_2(U)&=h^0+3{\rm i}h_\ast U -3h^\ast U^2-{\rm i}h_0 U^3,\label{eq:p2}
\end{align}
where $f$'s and $h$'s refer to RR and NS-NS fluxes. See Appendix \ref{sec:notation} for more details on this notation.
Being a no-scale superpotential for the K\"ahler modulus, the minimum for the scalar potential ${\cal V}$, supersymmetric or not,  is constrained to be positive or null. Since the dependence of ${\cal V}$ on  the K\"ahler modulus $T=\tau+{\rm i}\theta$ only comes from the K\"ahler potential $K$ 
\eq{
K=-3{\rm log}(U+U^*)-{\rm log} (S+S^*)-3{\rm log}(T+T^*),
}
it contains a flat direction on $\theta$ and a runaway direction on $\tau$. The moduli vacuum expectation values $U_0$ and $S_0$ are fixed by the equations $\partial_U {\cal V}=\partial_S {\cal V}=0$.  Since the potential is of the form
\eq{
{\cal V}=\frac{1}{8\tau^3}{\cal F}(U_0, S_0),
}
with ${\cal F}(U_0,S_0)$  a positive real function. The canonically normalized field $t_1=\sqrt{\frac{3}{2}}\ln(T+\bar T)$ serves to obtain 
\begin{eqnarray}
|\nabla {\cal V}|&=&\partial_{t_1}V=\sqrt{2G^{T\bar T}\partial_T V\partial_{\bar T}V}=\sqrt{ \frac{G^{T\bar T}}{2}}\partial_{\tau}V \nonumber\\
&=&\sqrt{6}|{\cal V}|.
\end{eqnarray}
and therefore, the first refined-dS criterium ($|\nabla {\cal V}|\geq c{\cal V}$) is fulfilled, notice that a SUSY solution implies ${\cal F}=0$. Similarly, since we have a flat direction on $\theta$. For the second derivatives we have:  
\eq{
\text{min}({\cal V}_{ij})=\rm{min} 
\begin{pmatrix}
    6{\cal F}\exp{(-\sqrt{6}t_1)}&0\\
	0&0
           \end{pmatrix}=6 \mathcal{V}
}          
and therefore the second dS criterion ($\text{min}({\cal V}_{ij})\leq -c' {\cal V}$) is violated. However it is necessary to restrict the range of values for $\tau$ in modulus space according to the Distance Conjecture (DC). This simple no-scale model is out of the Swampland for every value of $\tau$ subject to an upper bound.\\

\subsection{Toy model 2: non-SUSY \`a la KKLT}

 In this Section we review the scalar potential for a type IIB string theory toroidal compactification. The
superpotential dependence on the K\"ahler modulus arises from D7 branes.
This model is also a case that satisfies the refined Swampland constraints.\\

Let us consider a toroidal compactification with a superpotential given by
\eq{
W(U,S,T)= {\cal W}(U,S)+\widetilde{W}(T),
}
where the contribution depending on $T$ coming from gaugino condensation or instanton contributions from $D7$-branes as in KKLT, is given by $\widetilde{W}(T)=Ae^{-\lambda T}$ with $\lambda>0$.  Assume as well a hierarchy on the moduli such that $U$ and $S$ are fixed  independently of $T$. This is obtained by solving the equations $D_U{\cal W}=D_S{\cal W}=0$.  Since $D_U\widetilde{W}$ and $D_S\widetilde{W}$ are exponentially suppressed by $\tau$, the hierarchy assumption is valid and 
$D_U\widetilde{W}=D_S\widetilde{W}\sim 0$ for large values of $\tau$. In this scenario the scalar potential is given by
\eq{
V(\tau)=\frac{M_{pl}^4 \lambda}{2^7 \pi s_0 u_0^3}\left( |A|^2\left( \frac{1}{\tau^2}+\frac{\lambda}{3\tau} \right) e^{-2\lambda\tau}+\frac{ \gamma(\theta_0)}{\tau^2}e^{-\lambda\tau} \right),
}
with $\gamma(\theta)= \rm{Re}(\bar{A}{\cal W}_0e^{i\lambda\theta})$ and $\theta_0={\rm arg}(W_0)$ with ${\cal W}_0={\cal W}(U_0, S_0)$. The potential, as known, exhibits an AdS vacuum or a run-away direction on $\tau$ depending to the values on the involved constants. In the case of a runaway $\tau$-direction it has been pointed out in \cite{Ooguri:2018wrx} that the distance conjecture allows the potential to fulfill de refined dS criteria since large values on moduli space for $\tau$ would imply the apperance of extra light modes. The effective model is then consistent for large values of $\tau$, with values limited by the
appearance of extra light KK modes, such that an upper bound on $\tau$ must be imposed by hand.\\

Now,  as stated in \cite{Conlon:2018eyr} corrections to the superpotential on the K\"ahler modulus adds extra terms on the scalar potential. According to the refined dS conjecture,  whether these corrections lead us to the swampland or not, will be an indicative of compatibility with  a quantum gravity theory such as string theory.  In that context we shall explore under which conditions the inclusion of non-geometric fluxes satisfies the above bounds. This implies considering only tree-level corrections of the superpotential trough the inclusion of a linear term $\widetilde{W}(U_0, T)={\rm i}TP_3(U_0)$ in the superpotential which introduces an interaction of the K\"ahler modulus with the complex structure modulus.\\

%% file: modulih.tex
We consider type IIB string theory compactified on an isotropic six-dimensional torus with fluxes and orientifold 3-planes. The corresponding construction involves 6 real moduli fields and 8 different integer fluxes\footnote{As we shall comment, considering integer non-geometric fluxes seems to be a natural assumption.}. 
The corresponding superpotential reads\footnote{One sees from the expression of $W$ that there is a symmetry between moduli $S$ and $T$. This symmetry has been used to obtain flat directions which can be unflattened by precisely breaking the symmetry upon inclusion of new fluxes denoted as $P$-fluxes \cite{Aldazabal:2006up}. }
\begin{align}
W&={\cal W}(U,S)+ {\rm i}TP_3(U),\nonumber\\
&=P_1(U) -{\rm i} S P_2(U)+{\rm i} T P_3(U),
\end{align}
with $P_3$ being also a cubic polynomial on $U$ given by
\begin{align}
P_3(U)&=3 \left( b^0 + {\rm i}(2 b_*+\beta_*) U -( 2 b^*+\beta^*) U^2 - {\rm i} b_0 U^3 \right), \label{eq:p3}
\end{align}
with  $b$'s and $\beta$'s corresponding to non-geometric fluxes (see Appendix B for notation). The scalar potential $V$ depends on all moduli $U,S, T$ and some extrema are expected in a general flux configuration. \\

However we are interested in studying effective models with runaway directions, particularly on the K\"ahler field $\tau$. For that we present an analytical solution for fixing the vevs of $U$ and $S$ such that a kind of ``no-scale" behavior is present in the effective model. The first assumption for such a goal is the presence of hierarchies on the moduli. Therefore we proceed to clearly describe this assumption.\\

\subsection{On the Moduli Hierarchy Assumption}

In this Subsection a moduli hierarchy is discussed in the scalar potential. We write conditions to obtain a separation between the scales
of certain moduli.  Based on the above toy models, an interesting scenario emerges, where the complex-structure moduli and the axio-dilaton are stabilized in a first step and the K\"ahler 
moduli are stabilized in a second step.\\

Consider a superpotential $W=W(\phi_I)$ depending on $N$ moduli fields $\phi_I$ with $I=1\dots N$  and  let us assume that the vacuum expectation values for
some of the moduli fields $\phi_a$ are (almost) fixed independently of the rest of the moduli denoted as $\phi_i$ with $i\ne a$.
 We shall refer to this assumption as the {\it Moduli Hierarchy Assumption}. Under this scheme 
the vacuum expectation values of  $\phi_a$ are assumed to be barely modified by the dynamics of the rest of moduli $\phi_i$, implying strong constraints on flux configurations as we shall see. 
In terms of the scalar potential, the moduli hierarchy assumption implies that the fields $\phi_i$ are fixed at a minimum of the potential
\eq{
V((\phi_0)_a,\phi_{i})= \frac{M^4_{pl}}{4\pi}e^K\left(\sum_{i\ne a}D_i W D_{\bar{j}}\bar{W}K^{i\bar{j}}-3|W|^2\right)_{\phi_a=(\phi_0)_a},
}
where the fields $\phi_a$ are fixed at their vevs denoted $(\phi_0)_a$.\\

The implications of the assumption on the hierarchy of moduli strongly depend on the form of the superpotential. In this case we are thinking of a superpotential $W$ consisting of a component ${\cal W}$ which depends on the moduli $\phi_a$ and a second component $\widetilde{W}$ being a function of $\phi_i$ and containing interactions between $\phi_a$ and $\phi_i$. Those would be obtained by  compactification or dimensional reduction of string theory.
The complete superpotential is of the form
\eq{
W(\phi_a, \phi_i)= {\cal W}(\phi_a) + \widetilde{W}(\phi_a, \phi_i).
}

To clearly specify the constraints followed by our assumption it is important to observe that the scalar potential can be written as \cite{Kallosh:2014oja}
\eq{
V(\phi_a, \phi_i)= \frac{M^4_{Pl}}{4\pi}\left({\cal V}+\widetilde{V} + V_{int}\right),
}
where 
\begin{eqnarray}
{\cal V}&=& e^K\left(|D_I{\cal W}|^2K^{I\bar{I}}-3|{\cal W}|^2\right),\nonumber\\
\widetilde{V}&=&e^K\left(|D_i\widetilde{W}|^2 K^{i\bar{i}}-3|\widetilde{W}|^2\right),\nonumber\\
V_{int}&=&e^K\left( |D_a\widetilde{W}|^2 K^{a\bar{a}}+2\text{Re}\left((\phi_a) \text{Re}(D_a{\cal W}\cdot \widetilde{W}^\ast)  + (\phi_i) \text{Re}(D_i\widetilde{W}\cdot {\cal W}^\ast)-3 (\widetilde{W}{\cal W}^\ast)\right)\right).\nonumber\\
&&
\label{Eq:potenciales}
\end{eqnarray}
According to the anzatz, $(\phi_0)_a$ is a solution of $\partial_a {\cal V}|_{(\phi_0)_a}=0$ (see Appendix A for the specific case of the isotropic torus). \\

On the other hand, the vacuum expectation values for the moduli fields $\phi_i$, denoted $(\phi_0)_i$ should be determined by the system
\eq{
\partial_i V((\phi_0)_a, \phi_j)\big{|}_{\phi_i=(\phi_0)_i}=0,
\label{Eq:aprox1}
}
for each $i$. The assumed hierarchy for the fields $\phi_a$ would be consistent with the above method of computing the vevs and mass of the rest of the moduli if
\eq{
\partial_a(\widetilde{V}+ V_{int})\sim 0.
\label{Eq:aprox2}
}
The moduli mass  are given,  as usual,  by the eigenvalues of the mass matrix $M^2_{IJ}=\partial_{IJ}V$.
However, according to the assumed hierarchy,   the following constraints must be fulfilled
\begin{eqnarray}
\partial_{ab}(\widetilde{V}+V_{int} )&\sim& 0,\nonumber\\
\partial_{ai} V&\sim& 0.
\end{eqnarray}
such that
\eq{
M_{IJ}^2=
\begin{pmatrix}
m^2_{ab}&{m}^2_{aj}\\
{m}^2_{ib}& m^2_{ij}
\end{pmatrix}\sim
\begin{pmatrix}
\partial_a\partial_b{\cal V} & 0\\
0&\partial_i\partial_j V
\end{pmatrix}_{\text{min}},
}
with the corresponding eigenvalues $M_a^2$ and $M^2_i$ satisfying
\eq{
\frac{M^2_i}{M^2_a}< 1.
}
A mass hierarchy of order 10 can be obtained from F-terms in the presence of fluxes as described in the flux-scaling scenario \cite{Blumenhagen:2015kja}.  Our goal is to elucidate how the hierarchy moduli assumption determines whether an effective theory is in or out the Swampland.


%% file: hierarchys.tex
As stated before, we shall consider the GVW superpotential with tree-level corrections depending on the K\"ahler modulus. For that we assume that the complex structure modulus $U$ and the dilaton $S$ are fixed independently of $T$ implying that the contribution of non-geometric fluxes is sub-leading. In the following we shall use of the notation introduced in the last section. To start with, we consider the GVW superpotential written in the form 
$\mathcal{W}=P_1(U) -{\rm i} S P_2(U)$. Then we have a no-scale scalar potential of the form 
\eq{
{\cal V}=e^K(K^{SS}|D_S {\cal W}|^2+K^{UU}|D_U {\cal W}|^2)\,.
}
Fixing $U$ independently of $T$ implies finding a solution of $\partial_U {\cal V}=0$ which involves a dependence on $S$. If we assume that $U=U_0$ is also fixed independently of $S$, the solution of $\partial_U{\cal V}=0$ is the same as the equation $D_U{\cal W}=0$ (see Appendix A were we derive extremum scalar potential conditions with the covariant derivatives of the superpotential). We are therefore considering a model in which
\begin{align}
\begin{split}
D_S{\cal W}&=0, \\
D_U{\cal W}&=0.
\end{split}
\end{align}
Being a no-scale model, the last assertion also implies that at the minimum of the potential is at ${\cal V}_0=0$ where supersymmetry could or could not be broken by $T$. However, it is straightforward to see that a supersymmetric solution leads us to trivial solutions for $U_0$. Henceforth we consider the case in which $D_T{\cal W} \ne 0$. \\

Generically we know that a solution to the above equations implies turning on a $G_3$ form of type $(2,1)$ and $(0,3)$ \cite{Giddings:2001yu}. However, here we are interested in expressing the fluxes in terms of their symplectic components $(f_I, f^I, h_I, h^I)$. Hence, stabilization of $S$ is obtained from $D_S{\cal W}=0$ from which we obtain
\eq{\label{eq:sigma}
S_0=-{\rm i}\frac{P^\ast_1(U_0)}{P_2^\ast(U_0)}.
}
In this scheme the stabilized value of $S$ depends on  $U$. \\

\subsection{A particular solution}

In this Subsection we present an analytic solution to the conditions $D_S{\cal W}=D_U{\cal W}=0$.
Only RR and NSNS fluxes are on, generating a superpotential ${\cal W}(U,S)$ giving a setup
where complex structure and axio-dilaton are stabilized in a first step. As we shall see, a consistent solution
 will require a relation between fluxes.\\

A generic solution would imply to substitute the dilaton as a function of the complex structure in $D_U{\cal W}=0$. From this last equation
\eq{\label{eq:DUmink}
\left(P_1'(U)-\frac{3}{U+U^\ast}P_1(U)\right)-{\rm i}S\left(P_2'(U)-\frac{3}{U+U^\ast}P_2(U)\right)=0,\
}
with $P_1(U)$ and $P_2(U)$ different from zero in order to have ${\cal W}\ne 0$ at the minimum of the potential and $P'_i(U)$ being the derivative of $P_i$ with respect to $U$. A general analytical solution seems difficult to obtain since we also have to satisfy the tadpole condition. We proceed to consider a particular solution by solving (\ref{eq:DUmink}) written as:
\eq{\label{eq:subsoln}
(U+U^\ast)D_UP_i=P_i'(U)(U+U^\ast)-3P_i(U)=0, \qquad \text{for $i=1,2$.}
}
A common solution for $U$ in the above pair of equations will also stabilize the value of the dilaton $S$.  Indeed, there is a common root for the equations $(U+U^\ast)D_UP_1(U)=0$ and $(U+U^\ast)D_UP_2(U)=0$ if there is a relation among RR and NS-NS fluxes. 
 In order to see that, observe that
the above equations are also represented by
a quadratic polynomial on $\Re(U)$  and cubic in $\Im(U)$. As solutions we have\footnote{One additional solution has $\Re(U_0)=0$. We shall not consider such unphysical case.} 
\eq{
U_0=\frac{1}{2C_i}\left(\pm \delta_i+{\rm i}B_i\right),
\label{U_0}
}
for $i=1, 2$, where the RR fluxes $A_1,B_1,C_1$ and the NS-NS fluxes $A_2, B_2, C_2$ are given by
\begin{align}
\begin{split}
A_1&=f^0f^\ast-f_\ast^2,~\qquad  A_2= h^0h^\ast -h_\ast^2\\
B_1&=f_0f^0-f_\ast f^\ast \qquad B_2= h_0h^0-h_\ast h^\ast,  \\
C_1&=f_0f_\ast -(f^\ast)^2 \qquad C_2= h_0h_\ast-(h^\ast)^2 ,
\end{split}
\end{align}
and
\eq{
\delta_{1,2}&= \sqrt{4A_{1,2}C_{1,2}-B_{1,2}^2}.
}
We observe that one way for both equations to be simultaneously fulfilled, the flux coefficients must satisfy the relations
\eq{\label{eq:fluxrelations}
\frac{A_1}{C_1}=\frac{A_2}{C_2}\,\quad\quad 
\frac{B_1}{C_1}=\frac{B_2}{C_2},
}
while assuring a non-zero tadpole contribution from fluxes $f$ and $h$\footnote{These constraints together with Bianchi identities allow the tadpole for D7-brane charge to vanish only by flux contributions. This implies that no $D7$-branes or $O7$-planes are present in our model.}.  Notice that for both solutions the magnitude of $U_0$ is independent of fluxes $B$.
The value of $P_1$ at the minimum reads
\begin{eqnarray}
P_1(U_0)&=\frac{\delta_1^2}{2C_1^3}\big((f_0B_1-2f^\ast C)\mp {\rm i}f_0\delta_1\big),\nonumber\\
P_2(U_0)&=\frac{\delta_2^2}{2C_2^3}\left((h_0B_2-2h^\ast C_2)\mp {\rm i}h_0\delta_2\right).
\end{eqnarray}
In terms of the above fluxes, the dilaton vev is given by
\eq{
S_0=-{\rm i}\frac{\big[ (f_0B_2-2f^\ast C_2)\mp{\rm  i} f_0\delta_2\big]\big[(h_0B_2-2h^\ast C_2)\pm {\rm i}h_0\delta_2\big]}{(h_0B_2-2h^\ast C_2)^2+h_0^2\delta_2^2}, \label{S0eq}
}
from which the string coupling $s_0=e^{-\phi}=1/g_s$ reads
\eq{
s_0=\frac{(\pm 2 \delta_2C_2)(h_0f^\ast-f_0h^\ast)}{(h_0B_2-2h^\ast C_2)^2+h_0^2\delta_2^2}.
}
Using the tadpole condition (\ref{tadpole}) and the constraints (\ref{eq:fluxrelations}),  $s_0$ reduces to
\eq{
s_0 = \frac{2^4}{(-4 h_0 h_\ast^3-h_0^2 (h^0)^2+6 h_0 h_\ast h^0 h^\ast+3 h_\ast^2 (h^\ast)^2-4 h^0 (h^\ast)^3)^{1/2}}.
\label{s0}
}
Therefore, any physical solution implies a new flux constraint of the form
\eq{
\left( -4 h_0 h_\ast^3-h_0^2 (h^0)^2+6 h_0 h_\ast h^0 h^\ast+3 h_\ast^2 (h^\ast)^2-4 h^0 (h^\ast)^3 \right)<2^8.
}
Some comments are in order. First of all observe that $s_0$ only depends on 4 NS-NS fluxes $h$. In principle we start with 8 fluxes $h$ and $f$, but only 5 of them are free once we take the flux constraints \eqref{eq:fluxrelations} together with the tadpole cancellation condition. In that context we select 4 NS-NS fluxes $h$ and one RR $f$ corresponding to the 5 degrees of freedom. Second, since not all terms are positive in the denominator, it is possible to have some flux configurations leading us outside the physical region for which $s_0$ is smaller than unity.  \\

We have fixed $U$ and $S$ independently of the K\"ahler modulus $T$, which has to be incorporated. Therefore our next step is to consider such tree level correction on the superpotential, looking for the required conditions such that our hierarchy assumption is consistent.\\

\subsection{Non-geometric fluxes}

 In this Subsection we discuss the implications of the solutions for $U$
and $S$ presented in (\ref{U_0}) and (\ref{S0eq}) for RR and NSNS fluxes satisfying (\ref{eq:fluxrelations}). Now we incorporate non-geometric 
fluxes , such that the superpotential has a dependence on the K\"ahler modulus.

Consider now the whole superpotential of the form 
\eq{ 
W(U,S,T)= {\cal W}(U,S)+iTP_3(U),
}
with a scalar potential $V(T)=V(U_0, S_0, T)$. The values $U_0, S_0$ are the  previously computed vevs\footnote{Take notice of our notation in agreement with section 2.}. 
However, these values turn out to constraint the polynomial $P_3$ since  a root $U_0$ of the polynomial $(U+U^\ast)P_2'(U)-3P_2(U)=0$  is also a root of $P_3(U)$ for any set of fluxes on the isotropic torus.\\

This follows from the use of Jacobi Identities for the non-geometric fluxes, $Q\cdot H=Q\cdot Q=0$ from which it is possible to establish a set of relations among non-geometric and NS-NS fluxes. Before discussing the implications, let us first show that indeed $P_3(U_0)=0$. For the isotropic case there is a  particular solution for Jacobi (\ref{Jacobi}) and Bianchi identities (\ref{QQ})   given by
\begin{align}
\begin{split}
b^0&=\frac{A_2}{C_2}b^*\,,\\
b_0&=\frac{C_2}{A_2}b_*\,,\\
\beta_{*} &=\frac{B_2}{C_2} b^\ast-b_\ast\,,\\
\beta^{*} &= \frac{B_2}{A_2}b_\ast-b^\ast,
\end{split}\label{solution}
\end{align}
allowing us to express four non-geometric fluxes in terms of just two of them, namely $b_\ast$ and $b^\ast$. Notice the difficulty on having integer fluxes satisfying the above constraints. Then the polynomial $P_3$ takes the form
\begin{equation}
P_3(U)= Q(U) q_3(U),
\end{equation}
with
\begin{eqnarray}
Q(U)&=3(\frac{b^\ast}{C_2}+i\frac{b_\ast U}{A}),\nonumber\\
q_3(U)&= A_2 +iB_2 U -C_2 U^2.
\end{eqnarray}
We observe that $P_3$ depends only on  2 non-geometric fluxes through $Q(U)$ while $q_3(U_0)=0$.
Observe that $U_0$ depends only on NS-NS fluxes and in fact, it is the only solution of $(U+U^\ast)P_2'(U)-3P_2(U)=0$ with ${\Re}(U_0)\ne 0$. \\

Here we have shown that $P_3(U_0)=0$, now let us discuss the physical implications of our solution.\\

\subsection{Physical viability}

Once we have shown that $P_3(U_0)=0$ we must check if our solution is compatible with the moduli hierarchy assumption and proceed to study the implications on the scalar potential properties. First, notice from (\ref{Eq:potenciales})  that the scalar potential is  given by
\eq{
V(U_0, S_0, T)= \frac{M_{pl}^4}{4\pi}\frac{|P_3'(U_0)|^2}{3\cdot 2^5 u_0 s_0}\left(\frac{1}{\tau}+\frac{\theta^2}{\tau^3}\right),
\label{Eq:V(T)}
}
which is actually $V_{int}$ since ${\cal V}=\widetilde{V}=0$.  Hence we see that $V(T)$ reaches  a minimum at $\theta_0=0$ in the $\theta$-direction and  has a runaway direction on $\tau$.
Since $P_3(U_0)=0$ we also have
\begin{align}
D_U W&= D_U(iTP_3)=iTP'_3(U_0),\\
D_SW&=0,\\
D_TW&= -3\,\frac{\text{Im}\, \left( P_1 (U_0) P_2^\ast(U_0) \right)}{\tau P_2^\ast (U_0 )}.
\end{align}
The covariant derivatives are evaluated on the vevs $U_0$ and $S_0$. 
From these expressions we see that in order for the approximations  (\ref{Eq:aprox2}) to be valid and the moduli fields $U$ and $S$ to be fixed independently of $T$ it is necessary that the above vevs are not affected by the potential $V_{int}$, i.e. that $\partial_{S,U} V_{int}(U_0,S_0)\sim 0$. One way to fulfill this requirement is to restrict $V_{int}(U_0,S_)\sim 0$ and that $D_UW\sim 0$ in comparison with $\mathcal{V}$. This can be accomplished if $P'_3(U_0)$ vanishes or if 
\eq{
\frac{|TP_3'(U_0)|^2}{\tau^3}\ll 1,
\label{c1}
}
together with 
\eq{
|TP_3'(U_0)| \ll 1,
\label{c2}
}
which guarantee that $U$'s and $S$'s vev's are approximately kept at the values $U_0$ and $S_0$ respectively. Therefore our task now is to assure the viability of vanishing of $P_3'(U_0)$ or the above two constraints. Let us start by checking wether $P_3'(U_0)$ can vanish or not. First of all, in terms of fluxes
\eq{
|P_3'(U_0)|^2=9\left(\frac{\delta^2_2}{A_2C^2_2}\right)(A_2(b^\ast)^2-B_2b^\ast b_\ast +C_2b_\ast^2).
}
Vanishing of $|P_3'(U_0)|$ implies either that $\delta_2=0$, or that
\eq{
b^\ast =\frac{1}{2A_2} \left(B_2\pm {\rm i} \delta_2\right) b_\ast.
}
However, non-geometric fluxes $b_\ast$ and $b^\ast$ are real while $\delta_2$, besides being real, must also be different from zero (see (\ref{U_0})), implying that $|P_3'(U_0)|$ cannot vanish. Therefore the only option to be consistent with our hierarchy assumption is to tune on the non-geometric fluxes such that constraints (\ref{c1}) and (\ref{c2}) hold. As we shall see, this is deeply connected to the Swampland Distance Conjecture.\\

\subsubsection{Implications on the flux scaling-scenario}

Before discussing our model's consistency with the moduli hierarchy assumption  and the fixing of the cut-off scale by a proper selection of non-geometric fluxes, it is important to analyze the implications on the flux-scale scenario \cite{Blumenhagen:2015kja}. We will analyze the hierarchy of physical scales. 

The following hierarchy of scales is expected
\eq{
M_{pl}>M_s>M_{KK}>M_{U,S}>M_{T}.
\label{hierarchy}
}
$M_{pl},M_s,M_{KK},M_{U,S,T}$ denote the Planck-,  string-, Kaluza-Klein-
and moduli masses- scales respectively. Following conventions in \cite{Blumenhagen:2015kja}  we have
\begin{align}
M_s=&\frac{\sqrt{\pi}M_{pl}}{s_0^{1/4}(\mathbf{V})^{1/2}}=\frac{\sqrt{\pi}M_{pl}}{2^{3/4}(s_0\tau^3)^{1/4}},\nonumber\\
M_{KK}=&\frac{M_{pl}}{(4\pi)^{1/4}\mathbf{V}^{2/3}}= \frac{M_{pl}}{2(4\pi)^{1/4}}\frac{1}{\tau},\nonumber\\
m_{3/2}^2=&\frac{M^2_{pl}}{4\pi}e^{K_0}|W_0|^2=\frac{M_{pl}^2}{4\pi 2^5 s_0u_0^3}\frac{({\rm{Im}} (P_1(U_0)P_2^\ast(U_0)))^2}{|P_2(U_0)|^2}\frac{1}{\tau^3},
\end{align}
where $m_{3/2}$ is the gravitino mass and $\mathbf{V}$ is the volume of $T^6$ in the Einstein frame. We have used the relation
\eq{ 
-2\rm{log}(\mathbf{V})=-3\rm{log}(2\tau).
}
Since $s_0, \tau>2$ for a supergravity description to be valid in a physical region, it follows that $M_{pl}>M_s>M_{KK}$ in concordance with (\ref{hierarchy}). Notice that all mass scales as well as the moduli masses are unfixed and depend inversely on $\tau$,
while the relevant ratios are determined by
\eq{
\frac{M_s}{M_{KK}}= 2\pi\left(\frac{2\tau}{s_0}\right)^{1/4}.
}
Observe that this behavior is the same as the models with frozen complex structure studied in the flux-scaling scenario \cite{Blumenhagen:2015kja}, implying that $\tau>s_0$. Therefore, for consistency and taking $\theta=\theta_0=0$, we get
\eq{
\tau > {\rm{max}}\left(|P_3'(U_0)|^2, s_0\right).
}
The scale of supersymmetry breaking is determined by the non-vanishing F-term
\eq{
F^T=e^{K/2}K^{TT^\ast}(D_TW)^\ast,
}
evaluated at $S_0$ and $U_0$ and given by 
\eq{
M_{pl}/m_{SUSY}\sim \frac{s_0 u_0|P_2(U_0)|}{{\rm Im}(P_1(U_0)P_2^\ast(U_0))} \, .
}
Notice that the scale at which SUSY is broken is determined by fluxes $f$ and $h$ where non-geometric  ones are not playing a role since the above ratio does not depend on $\tau$. Finally, the moduli mass eigenvalues depend on $\tau$ as 
\eq{
M_i\sim \frac{1}{\tau^3},
}
following that 
\eq{
\frac{M_{KK}}{M_i}\sim \tau^2,
}
where $i=U,S,T$. Since $\tau$ is not fixed, there is only a range of values in moduli space in which $M_{KK}>M_i$, actually for $\tau>1$. It is then  important to check  the bounds for $\tau$.

\subsection{Moduli Hierarchy and the Swampland Distance Conjecture}

Up to here we have presented a model in which the presence of non-geometric fluxes have not altered the runaway profile of the scalar potential on the $\tau$-direction but have stabilized $\theta$. Since any scalar potential with a dependence on $\tau^n$ for any integer value of $n$ satisfies
\eq{
|\nabla V|=\sqrt{\frac{2}{3}}n V, \,.
}
it automatically satisfies one of the  refined dS bounds. The same occurs to the potential (\ref{Eq:V(T)}) even for $\theta\ne 0$ since
\eq{
|\nabla V|=\sqrt{\frac{2}{3}}\left(V+ \frac{2{\cal H}\theta^2}{\tau^3}\right),
}
with ${\cal H}= M_{pl}^2|P_3'(U_0)|^2/48 \pi u_0s_0$.
According to the refined dS conjecture such potentials can be considered to be out of the Swampland.  Therefore, the moduli hierarchy  assumption allows us to have a model with tree-level corrections on the superpotential depending on the K\"ahler modulus by the presence of non-geometric fluxes which is actually compatible with the refined dS criteria.  A priori there is no obstruction for this model to Such a model to be an effective theory compatible with a quantum gravity theory such as string theory. Notice however that despite of $S$ and $U$ being stabilized, all moduli masses still depend on $\tau$ for which they remain unfixed unless there is some criteria to constrain the value of $\tau$.\\

In the following we shall use the constraints on our moduli hierarchy assumption to derive some bounds on $\tau$. Even more we shall show that they are compatible with the distance conjecture, allowing us to establish a cutoff scale at which the effective model is valid.\\

From constraints (\ref{c1}) and (\ref{c2}), we obtain that at $\theta=0$
\eq{
|P_3'(U_0)|^2 \ll \tau \ll \frac{1}{|P_3'(U_0)|},
\label{tau}
}
which is an available range of $\tau$ if
\eq{
|P_3'(U_0)|\ll 1.
\label{rangeP3}
}
Notice that this is a restriction on non-geometric fluxes since NS-NS and RR fluxes have been already fixed at higher scales.  The above range of viable displacement on $\tau$ is a direct consequence of the  Moduli Hierarchy Assumption and allows us to estimate the range of scales at which our effective model is valid.  First of all, a supergravity description of the 10-dimensional model requires the internal volume $\mathbf{V}_{6D} >1$ with
\eq{
\mathbf{V}_{6D} =\mathbf{V} M_S^6,
} 
with $\mathbf{V}$ being the volume of $T^6$ in the Einstein frame. Therefore it follows that $\tau>1/2$ implying that $|P_3'(U_0)|<2$.\\

Second of all, from the bounded value for $\tau$ in (\ref{tau}) we have that the gravitino mass is constrained to the values
\eq{
\frac{M_{pl}^2}{4\pi 2^5 s_0u_0^3}\frac{({\rm{Im}} (P_1(U_0)P_2^\ast(U_0)))^2|P_3'(U_0)|^3}{|P_2(U_0)|^2}\ll m^2_{3/2}\ll \frac{M_{pl}^2}{4\pi 2^5 s_0u_0^3}\frac{({\rm{Im}} (P_1(U_0)P_2^\ast(U_0)))^2}{|P_2(U_0)|^2|P_3'(U_0)|^6}\,,}
indicating that for the gravitino mass to have an available range of values, $|P_3'(U_0)|$ must be less than unit.\\

A third important consequence of (\ref{tau}) is the following: it has been conjectured that moduli fields can not take large displacements otherwise massive fields interacting with the moduli must be taken into account. In such context and by taking string theory as the quantum gravity theory, the displacements are argued to be of the form \cite{Scalisi:2018eaz}
\eq{
\Delta \hat\tau< \frac{1}{\lambda}\log{\frac{M_S}{\Lambda_{SW}}}
}
where  $\hat\tau$ is the canonical normalized K\"ahler modulus and $\lambda$ has been typically taken of order 1\footnote{See Reference \cite{Scalisi:2018eaz} for a discussion on the scale of $\lambda$}. In our case, given the K\"ahler potential, $\lambda=2/\sqrt{3}$ and
\eq{
\hat\tau=\frac{\sqrt{3}}{2}\rm{log}(\tau),
}
implying that $\Lambda_{SW}$ is  fixed by $|P_3'(U_0)|$. Once  $f$ and $h$ fluxes have been chosen, it depends only on  non-geometric fluxes $b_\ast$ and $b^\ast$. Therefore
\eq{
\Lambda_{SW}\sim M_{S}|P_3'(U_0)|.
\label{LSW}
}
Notice that small values for $|P_3'(U_0)|$ would fix the scale $\Lambda_{SW}$ below string mass. This means that the canonical normalized field $\hat\tau$ have a non-zero range in which the model is consistent. Otherwise, for values of $|P_3'(U_0)|$  greater than unity, $\hat\tau$ has a zero range of consistent values.\\

\begin{figure}
\centering

\begin{tikzpicture}


\node[right] at (0.2,8) {$\tau$};
\node[right] at (10.5,0.5) {$| P_3' \left( U_0 \right)|$};
\node[right] at (0.1,4.2) {$\tau_0$};
\node[right] at (6,6) {Swampland};
\node[right] at (6,2) {Swampland};
\node[red,right] at (2.5,1.5) {Range of $\tau$};
\node[below] at (2,0) {$|P_3'|_{\text{min}}$};
\node[below] at (10,0) {$|P_3'|_{\text{max}}$};
\draw[blue,very thick] (2,1.5) -- (2,6.5);
\filldraw[black] (10,4) circle (3pt);
\draw[very thick] (2,6.5) to [out=330,in=170] (10,4);
\draw[very thick] (2,1.5) to [out=30,in=190] (10,4);
\draw[very thick] (2,5.5) to [out=330,in=170] (10,4);
\draw[very thick] (2,2.5) to [out=30,in=190] (10,4);
\path [fill=cyan] (2,5.5) to [out=330,in=170] (10,4) to [out=190,in=30] (2,2.5) -- (2,5.5);
\draw[red,very thick] (2.5,2.75) -- (2.5,5.26);
\draw[gray, very thick] (0,4) -- (10,4);
\draw[red,thick,->] (3.5,1.7) -- (2.5,3.7);
\draw[black,thick,->] (7.0,5.7) -- (4.5,4.7);
\draw[black,thick,->] (7.0,2.3) -- (4.5,3.3);
\draw[black, thick,->,line width=1mm] (0,-1) -- (0,8);
\draw[black, thick,->,line width=1mm] (-1,0) -- (12,0);
\draw[black,very thick] (2,-0.1) -- (2,0.1);
\draw[black,very thick] (10,-0.1) -- (10,0.1);
\end{tikzpicture}
\caption{Ranges of validity for $\tau$ in terms of the cut-off scale $|P_3'(U_0)|$. A wider range is obtained from small values of $|P_3'(U_0)|$ while for larger values,  $\tau$ is so constraint that the model is inconsistent entering into the Swampland before reaching $\tau=\tau_0$.}
\label{taurange}
\end{figure}
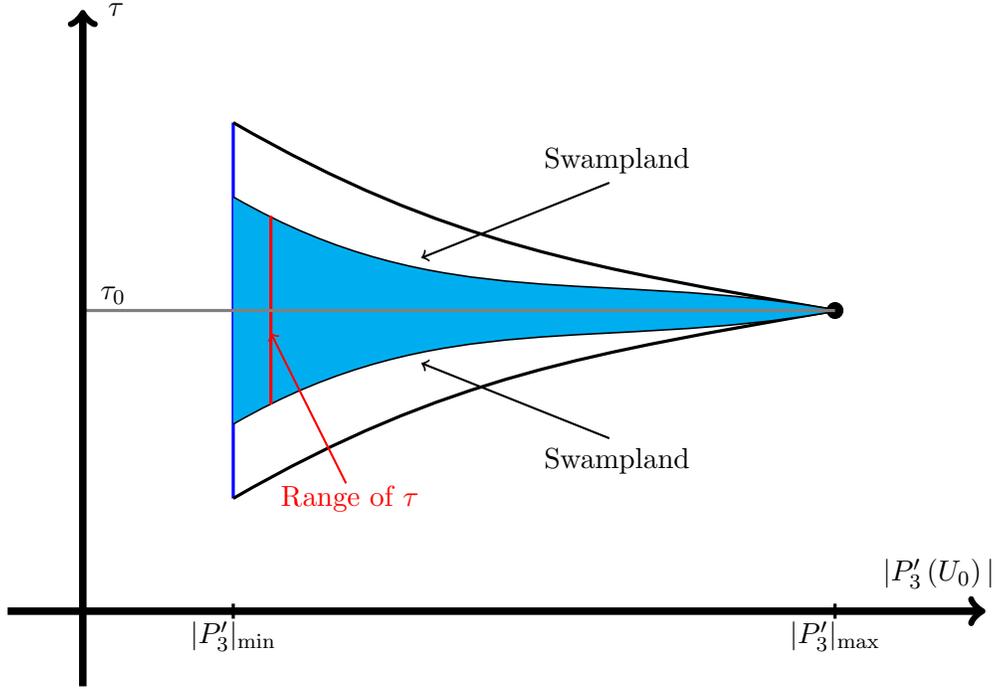

Therefore, by all the above implications, the range of viability for $\tau$ is fixed as\footnote{See \cite{Kachru:2018aqn} for comments on the establishment of viable ranges for moduli}
\eq{
{\rm{max}}\Big(\frac{|P_3'(U_0)|^{2}}{12 u_0s_0}, \frac{1}{2}\Big)\ll \tau \ll \frac{1}{|P_3'(U_0)|}.
}
Hence, the smaller the value for $|P_3'(U_0)|$, the larger the allowed range of displacement for $\tau$. For $|P_3'(U_0)|=\rm{max} (2, (12u_0s_0)^{1/3})= \tau_0$, $\tau$ in principle is fixed to a single value although the model is not consistent.  Notice that with large values for $|P_3'(U_0)|$ (which implies a better approximation consistent with the hierarchy assumption) the range for $\tau$ diminishes making more difficult to satisfy the  Swampland Distance Conjecture. However all we need is to have a non-zero range of viability for $\tau$ defined far away from the minimum and maximum values established by the hierarchy assumption while having $\tau>1$  for SUGRA to be a valid approach. See Figure \ref{taurange}. The question is if one can have such scenarios for concrete flux configurations.


\subsection{A numerical analysis}
\begin{figure}[htbp]
   \centering
   \includegraphics[width=12cm]{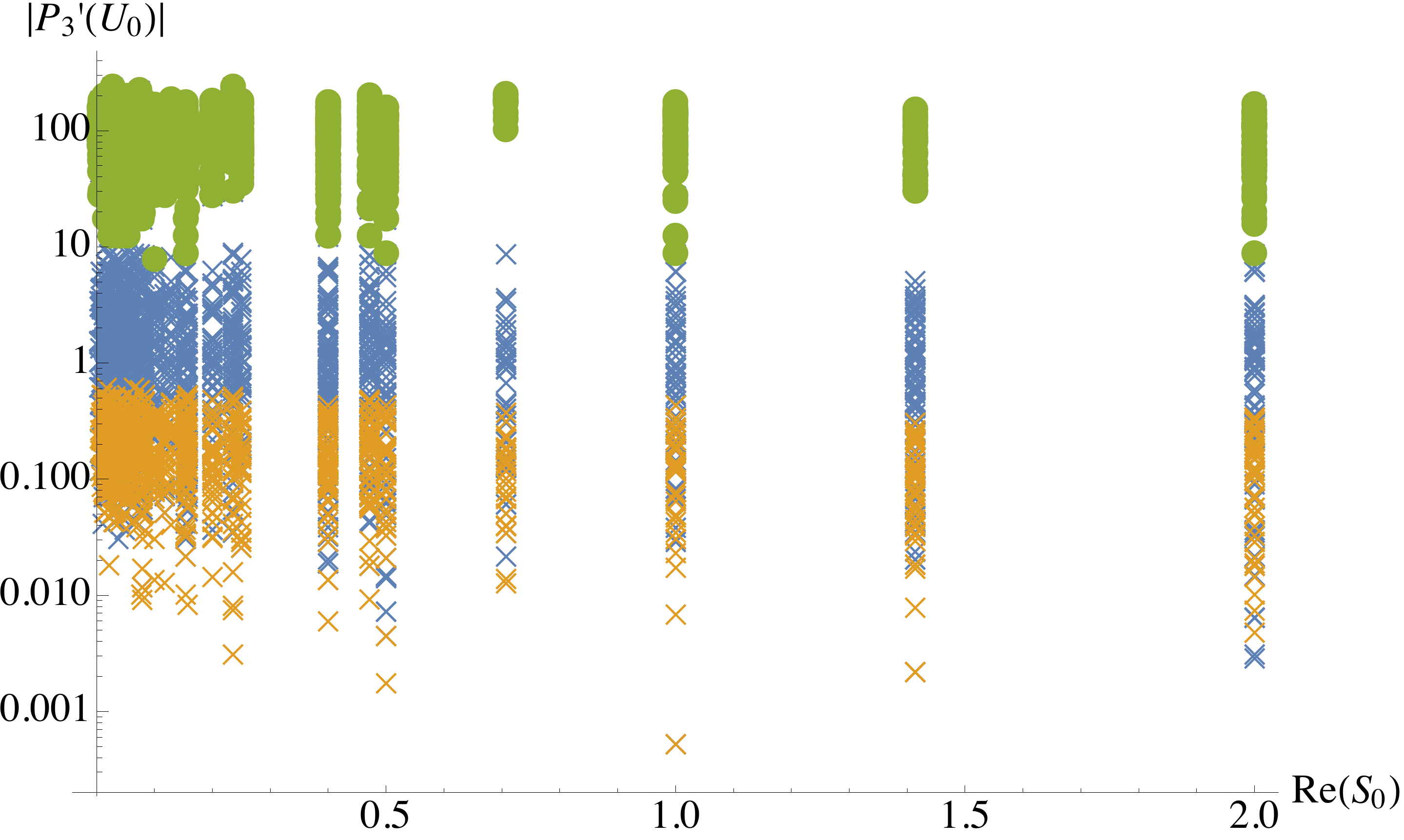} 
   \caption{Here we plot the values of $|P_3'(U_0)|$ against $s_0$ for eight thousand different flux configurations. Green dots indicate even-integer fluxes. Blue and orange crosses refer to fractional non-geometric fluxes. The former ones correspond to values between 1 and 0.1, while the later correspond to values between 0.1 and 0.01. Notice the discrete pattern of $s_0$.}
   \label{fig:P3p}
\end{figure}

From all the above it seems there exists a link between the Moduli Hierarchy Assumption and the distance conjecture, establishing a range of viability for $\tau$. Taking into account all the constraints, our model possesses 7 degrees of freedom (4 NS-NS, 1 R-R and 2 non-geometric fluxes) implying the necessity of  a numerical analysis. Using only even integer fluxes we find 8000 flux configurations fulfilling all restrictions. The results are plotted in Figure \ref{fig:P3p}.\\

 Interestingly, we find that for integer fluxes, none of the cases contains a small value for $|P_3'(U_0)|$ indicating that the model is not consistent. This means that the upper bound on $\tau$ is smaller than unity, destroying the supergravity approach. However, if one considers fractional non-geometric fluxes $b_\ast$ and $b^\ast$, the values for $|P_3'(U_0)|$ become less than one, and in turn it allows a consistent range for $\tau$.\\
 
 A particularly surprising issue that becomes evident from this plot is the discretness of the values of $s_0$ (which is independent of the non-geometric fluxes). The numerical evidence points out a maximum value for $s_0$ of 2 (although by considering odd-fluxes this can increase). This feature was also noticed in \cite{Junghans:2018gdb}
 and it is probably related to the high number of constraints (as shown in the expression (\ref{s0}) for $s_0$ in terms of fluxes which only depend on 4 NS-NS fluxes. Discrete values for the string coupling seem to be related to strong constraints after compactification, as fulfilling the tadpole condition or directly related to the topology of the internal space.\\
 
 Similarly, it is possible to have a  simple argument to scketch  how fractional fluxes are linked to small values of $|P_3'(U_0)|$. By assuming quantization of NS-NS and R-R fluxes one could assume its extension to non-geometry fluxes by imposing quantization on the action of non-geometric fluxes on $(p+1)$-forms, as
 \eq{
 \frac{1}{(2\pi)^{p-1}}\int_{\Sigma_p} Q\cdot \omega_{p+1} =n \in \mathbf{Z},
 }
 %
 %
 %
From (\ref{LSW}) it follows that
 \eq{
\frac{1}{(2\pi)^{p-1}} \int_{\Sigma_p} Q\cdot \omega_{p+1}=n\left( \frac{M_s|P_3'(U_0)|}{\Lambda_{SW} }\right)^{p-1}.
 } 
Hence,  for $|P_3'(U_0)|$ less  than one,  $|P_3'(U_0)|^{p-1}$ can be approximated as $1/k$ with  integer $k>1$. Therefore, up to $\Lambda_{SW}$ 
 \eq{
 \int_{\Sigma_p} Q\cdot \omega_{p+1}=\frac{M_s^{p-1}}{(\Lambda_{SW} )^{p-1}}\frac{n}{k}.
 }
 The $p$-form $Q\cdot \omega_{p+1}$ has fractional values in an effective theory  allowing for $b_\ast$ and $b^\ast$ to be fractional. The presence of non-integers and non-constant fluxes have been already considered in literature as fluxes sourcing punctures on a sphere \cite{ Candelas:2014kma, Candelas:2014jma, Damian:2016lvj} or fractional fluxes arising as a consequence of the  topology of the internal manifold \cite{Camara:2011jg, Garcia-Compean:2013sla}. Fractional fluxes can be considered as the result of the backreaction of the metric by the presence of non-geometric fluxes or equivalently by assuming T-duality on the internal manifold threaded with NS-NS fluxes.  Under this perspective, Dirac quantization is a feature compatible with string theory in its ten dimensional version which can be modified by an specific compactification setup. This is consistent with our theory as soon as the quantization is reinforced at high scales but weakened at lower energies. We have shown that this is indeed our case.

%

%% file: quintessence.tex
In this Subsection we particularize our solution to a set of fluxes satisfying (\ref{eq:fluxrelations}), Bianchi identities
and Tadpole cancellation conditions.  In this scheme we analyze the implications for the
Swampland constraints.

Let us focus on an example which satisfies all the constraints, namely{\color{red}:}
\begin{align}
b^0 = -\frac{b^\ast h^0}{h^\ast}\,, b_0 = - \frac{b_\ast h^\ast}{h^0} \,, \beta_\ast= b_\ast \,, \beta^\ast = - b^\ast \,, \\
f_\ast =  \frac{8}{h^\ast} \,, f^0 = \frac{f^\ast h^0}{h^\ast} \,, f_0 = \frac{8}{h^0} \,, h_\ast = h_0 = 0 \,,
\end{align}
thus one can see that in order to get even integer fluxes the NS fluxes are highly constrained. Indeed the only values allowed for the NS fluxes are $\pm$4 and $\pm 2$ in Planck units. The $U$ and $S$ moduli are fixed at 
\eq{
U = \left(-\frac{h^0}{h^\ast}\right)^{1/2}  \,,\quad S = \frac{2^3}{(-h^0 (h^\ast)^3)^{1/2} } +\frac{f^{\ast}}{h^\ast}  \text{i} \,,
}
which is a solution of the scaling type and a physical solution implies that $h^0$ and $h^\ast$ have opposite signs. Thus, in order to stay in the perturbative regime it is required that $|h^{\ast}| = 2$ (otherwise $s_0 < 0$) which is compatible with the flux quantization condition.  As is stated by \cite{Junghans:2018gdb},  it is possible to evade the Dine-Seiberg problem and to keep the theory in the perturbative regime just by fluxes, if the dilaton is stabilized at a value that it is not exponentially large.  The mass hierarchy is controlled by the value of $|P_3' (U_0)|$, which in terms of the fluxes is written as
\eq{
| P_3'(U_0) | = 6 \bigg| b_\ast + b^{\ast} \left(-\frac{h^0}{h^\ast} \right)^{1/2} \bigg|, 
}
and it has to be as small as possible. Now, since $h^0$ and $h^\ast$ have oposite signs, the magnitude of $P_3'(U_0)$ lies in a circle of radius of order $\mathcal{O} \left( 10 \right)$ violating the hierarchy condition. The hierarchies are preserved if we consider fractional non-geometric fluxes, which apparently violates the Dirac quantization condition. However, the cohomology group at which the non-geometric fluxes belong has to be determined. We let this subtle question for future work and we shall proceed with the approach of considering  non-geometric fluxes with magnitude less than 1, preserving a parametric control on the mass hierarchies. Fixing the values of the NS fluxes as $h^0 = -h^\ast = -2 $, the scalar potential takes the form
\eq{
V =\frac{3}{2^2}\frac{ b_\ast^2 + (b^\ast)^2 }{\tau}\,,
}
which corresponds to the potential for a quintessence scalar rolling to positive values. Since the quintessence field is represented by the K\"ahler modulus it can potentially lead to fifth-forces through its coupling with SM fields. However, since the rolling of the scalar field is parametrized by the non-geometric fluxes, it could be slow enough to effectively fix the couplings to SM fields avoiding fifth-forces. The flatness of the potential in such quintessence models have been recently explored \cite{PhysRevD.94.103526}.\\

This model breaks SUSY spontaneously though F-terms, with the sgoldsitno direction pointing mainly in the complex structure direction as $\tau$ becomes larger as
\eq{
F_i = \langle 2 \cdot 3( b_\ast - i b^\ast ) \tau , 0 , \frac{2^4\cdot3}{\tau} \text{i} \rangle \,, \quad i=U,S,T.
}
where by construction the $S$ direction is set to zero, and the swampland criteria 
\eq{
\lim_{\tau \rightarrow \infty} \frac{\nabla V}{V} = \frac{\sqrt{17}}{2} \,, 
}
is satisfied. As noticed in \cite{Damian:2018tlf}, this swampland criteria is not parametrically controlled by fluxes, instead it is possible to get a numerical control by a suitable choice of the scalar potential. The Moduli Hierarchy Assumption implies that Eq. (\ref{tau}) must hold, which for this particular solution can be written as
\eq{
b_\ast^2 + \bigg| \frac{h^0}{h^\ast} \bigg| \big(b^\ast)^2 <  \left( \frac{2}{3 h^\ast} \right)^{4/3} \,, 
}
thus, for $h^0 = -h^\ast$, the allowed non-geometric fluxes lie in a circle of radius  $\left( \frac{2}{3 h^\ast} \right)^{2/3}$ which is smaller than 1. Thus, together all the swampland criteria are satisfied if the non-geometric fluxes take fractional values less than 1. In this way the field range allowed by the distance conjecture can be parametrically controlled.

%% file: conclusions2.tex
In this work we studied  a compactification on an isotropic six-dimensional torus with non-geometric fluxes, orientifold 3-planes and no D-branes. Validity of T-duality in the effective four-dimensional theory $-$from which non-geometric fluxes have been introduced$-$ non-geometric flux quantization, extension of tadpole and Bianchi identities are assumed.
All of them constitute self-consistent assumptions inspired directly from string theory.  Notice that by assuming T-duality, interaction of non-geometric fluxes with K\"ahler moduli is introduced suggesting the existence of a shift-symmetry on the K\"ahler modulus derived from a symmetry on the non-geometric fluxes.\\

By a proper selection of fluxes it is possible to give a vev to the complex structure and the axio-dilaton moduli independently of the K\"ahler modulus and therefore independently of the choice of non-geometric fluxes. We called it the Moduli Hierarchy Assumption and an analytical solution to stabilize the complex structure and the axio-dilaton in this form is reported. For that we took a second assumption: a particular solution relating NS-NS and R-R fluxes  as shown in  Eqs.(\ref{eq:fluxrelations}). Together with the extension of Bianchi Identities and Tadpole conditions it is possible to show that such assumption leads us to two important consequences: First, that there are only 2 unconstrained non-geometric fluxes\footnote{The rest of the fluxes satisfy all constraints and are quantized. From the set of 8 NS-NS and R-R fluxes only 5 are free.} ($b_\ast$ and $b^\ast$).  Second, that the superpotential component depending on the K\"ahler moduli vanishes once evaluated at the vevs of $U$ and $S$. These two results restrict the effective model to have a runaway direction along the real part of the complex K\"ahler modulus $\tau$ in agreement with the Refined dS conjecture.\\

Also, we showed that for this particular model, the Distance Conjecture is fulfilled as a consequence of a particular selection of  flux configuration on which the hierarchy assumption on moduli is based. Particularly we found that $\tau$ is restricted  to have finite specific displacements for which infinite distances in field space would turn the effective theory inconsistent.  Moreover, all different scales, (depending on $\tau$) show a hierarchy as  expected in models such as the flux-scaling scenario. In this context, it is possible to compute the scale $\Lambda_{SW}$ at which the effective theory is valid turning out to be established solely by  two non-geometric fluxes $b_\ast$ and $b^\ast$. \\

After a numerical computation of near three thousand different configurations we were not able to find concrete examples with integer values for the non-geometric fluxes compatible with the Moduli Hierarchy Assumption, meaning that integer non-geometric fluxes does not allow a range of field displacement while fractional values establishes a physical consistent model.  We also noticed that due to tadpole constraints, the values for the string coupling shows a discrete pattern as suggested in \cite{Junghans:2018gdb}.\\

Fractional fluxes can arise at the compactification scale due to the internal manifold topology. In our case could be a consequence of the backreaction of the internal metric by assuming T-duality. This issue, although very well  known by the community, has been ignored in order to stabilize the internal volume and make estimations on the KK scale. Since the distance conjecture is deeply connected with the internal volume by having a runaway direction on $\tau$, it is expected to have definite range in the $\tau$ field space. We also illustrated how the fixing of the scale $\Lambda_{SW}$ permits us to argue that fractional values for non-geometric fluxes are expected.\\

In summary we have shown that the Moduli Hierarchy Assumption together with the presence of fractional non-geometric fluxes restrict the modulus $\tau$ to have finite distance displacements in the field space in agreement to the Distance Conjecture, while the effective cut-off scale is  completely fixed by non-geometric fluxes. Since the Hierarchy on moduli is constructed by a proper selection of NS-NS and RR fluxes our particular analytical solution for the moduli stabilization allowed us to connect the entire flux configuration with an effective model in which the run away direction is restricted to finite ranges. Infinite distances would imply the breakdown of the effective theory since the assumed hierarchy on moduli would not be fulfilled. It would be interesting to study the possible appearance of a tower of massless modes for large values of $\tau$ as expected. We leave this important task for future work.\\

Finally let us mention that we presented what we consider is a reliable method to construct effective models based on non-geometric fluxes. Along the way, we have elucidated the necessity to remove some assumptions, as the quantization of non-geometric fluxes, which allows to re-enter into the Landscape. If the Swampland criteria indeed divides the field theory space into two types of effective models, the requirements for  some of them to be in the Landscape could establish a way to understand implications of a quantum theory of gravity in four-dimensional effective theories.\\

%% file: appendix1.tex
The extrema of the scalar potential can be recast in terms of the covariant derivatives of the superpotential. In the simpler case
where only an $U$ dependence of the superpotential exists, one obtains that the condition $\partial_U V=0$
 is equivalent to $D_U W=0$. When only a dependence on $U,S$ is on the conditions
 $\partial_U V=\partial_S V=0$ can be satisfied simultaneously if $D_U W=D_S W=0$.
 There is however another solution, that is  $\partial_S V=0$ is satisfied by $D_U W=\partial_S D_UW=0$;
 and $\partial_U V=0$ is satisfied by $D_S W=\partial_U D_SW=0$. When
 the three moduli are on the conditions $\partial_U V=\partial_S V=\partial_T V=0$
 can be satisfied simultaneously for $D_U W=D_S W=D_T W=0$. But there
 are other simple cases which we summarize in the table \ref{extremosS}.
 \begin{table}[htp]
\caption{Particular extrema of the scalar potential in terms of superpotential covariant derivatives.}
\begin{center}
\begin{tabular}{|c|c|}
 \hline \hline
$\partial_U V=0$&$W(U)$ condition  \\ \hline 
 & $D_U W=0$  \\ \hline 
 \hline \hline
$\partial_U V=\partial_S V=0$& $W(U,S)$ condition  \\ \hline 
$\partial_U V=0$&  $D_U W=D_SW=0$ \\
 &  $D_U W=\partial_S D_UW=0$ \\
 $\partial_S V=0$&  $D_U W=D_SW=0$ \\
 &  $D_S W=\partial_U D_SW=0$ \\ \hline \hline
 $\partial_U V=\partial_S V=\partial_T V=0$&  $W(U,S,T)$ condition \\ \hline 
 $\partial_U V=0$&  $D_U W=D_SW=D_TW=0$ \\
  & $D_U W=\partial_T D_U W=0$ \\ \hline
 $\partial_S V=0$&  $D_U W=D_SW=D_TW=0$ \\
  & $D_S W=\partial_T D_S W=0$ \\ \hline
   $\partial_T V=0$&  $D_U W=D_SW=D_TW=0$ \\
  & $D_T W=\partial_U D_T W=0$ \\ \hline \hline
\end{tabular}
\end{center}
\label{extremosS}
\end{table}

%% file: appendix2.tex
We consider  a compactification on a six-dimensional torus in the presence of NS-NS and RR three-form fluxes. The corresponding superpotential is given by
\eq{
{\cal W}(U,S)=\int G_3\wedge \Omega,
}
where $G_3=F_3-iS H_3$. In terms of the 3-form cohomology symplectic basis $(\alpha^I, \beta _I)$, we have that
\begin{eqnarray}
F_3&=& f_I\alpha^I-f^I\beta_I,\nonumber\\
H_3&=&h_I\alpha^I-h^I\beta_I, 
\end{eqnarray}
with $G_3=F_3-iSH_3=g_I\alpha^I-g^I\beta_I$, where
\begin{eqnarray}
g_I&=&f_I-iS h_I,\nonumber\\
g^I&=&f^I-iS h^I.
\end{eqnarray}
The symplectic basis is given by
\begin{eqnarray}
\alpha^0&=& dx^1\wedge dx^2\wedge dx^3, \qquad \beta_0=dx^4\wedge dx^5\wedge dx^6, \nonumber\\
\alpha^1&=&dx^1\wedge dx^5\wedge dx^6, \qquad \beta_1=dx^4\wedge dx^2\wedge dx^3, \nonumber\\
\alpha^2&=&dx^4\wedge dx^2\wedge dx^6, \qquad \beta_2=dx^4\wedge dx^5\wedge dx^3, \nonumber\\
\alpha^3&=&dx^4\wedge dx^5\wedge dx^3, \qquad \beta_3=dx^1\wedge dx^2\wedge dx^6,
\end{eqnarray}
while flux components are 
\begin{eqnarray}
h_I&=& (h_0, h_1, h_2, h_3)=(H_{123}, H_{156}, H_{426}, H_{453}),\nonumber\\
h^I&=& (h^0, h^1, h^2, h^3)= (H_{456}, H_{423}, H_{453}, H_{126}),\nonumber\\
f_I&=& (f_0, f_1, f_2, f_3)=(F_{123}, F_{156}, F_{426}, F_{453}),\nonumber\\
f^I&=& (f^0, f^1, f^2, f^3)= (F_{456}, F_{423}, F_{453}, F_{126}).
\end{eqnarray}

In the considered model the complex structure is identical for the three tori $T^2$ and it is determined by the complex coordinate for each $T^2_i$, $z_i=x^i+iU y^i$. Thus the (3,0) holomorphic form reads
\begin{eqnarray}
\Omega&=&dz^1\wedge dz^2\wedge dz^3,\nonumber\\
&=& \alpha^0+iU(\beta_1+\beta_2+\beta_3)-U^2(\alpha^1+\alpha^2+\alpha^3)-iU^3 \beta_0.
\end{eqnarray}
We have considered a unique $U$ for the isotropic $T^6$ and use the following notation:
\begin{eqnarray}
\alpha^\ast&=&\alpha^1=\alpha^2=\alpha^3,\nonumber\\
\beta_\ast&=&\beta^1=\beta_2=\beta_3,\nonumber\\
f^\ast&=&f^1=f^2=f^3,\nonumber\\
f_\ast&=&f_1=f_2=f_3,\nonumber\\
h_\ast&=&h_1=h_2=h_3\nonumber\\
h^\ast&=&h^1=h^2=h^3.
\end{eqnarray}
from which the superpotential can be written as
\eq{
{\cal W}=g^0+3{\rm i}Ug_\ast-3g^\ast U^2-{\rm i}g_0U^3.
}
In terms of the NS-NS and RR fluxes, the superpotential reads
\eq{ \label{eq:super1}
{\cal W}= P_1(U)-{\rm i}SP_2(U),
}
where $P_i(U)$ are cubic polynomials on $U$ shown in expressions (\ref{eq:p1}) and (\ref{eq:p2}). Finally, for a compactification on an isotropic $T^6$ in the presence of $O3$-planes, the tadpole condition reads:
\eq{
\frac{1}{2}\int F_3\wedge H_3= {\cal N}_{O3},
}
where ${\cal N}_{O3}$ measures the contribution of an $O3^-$-plane to the internal $D3$-brane charge. In terms of the fluxes $f$ and $h$ the above expression reduces to
\eq{
f_0h^0+3f_\ast h^\ast-3f^\ast h_\ast -f^0h_0=16.
\label{tadpole}
}

\subsection{Superpotential with non-geometric fluxes}

Now we shall turn on non-geometric fluxes, meaning that we are considering a superpotential of the form \cite{Aldazabal:2006up}
\begin{eqnarray}
W(U,S)={\cal W}(U,S)+\frac{1}{\kappa^2}\int \left(Q\cdot J_c\right)\wedge \Omega,
\end{eqnarray}

with the 3-form $Q\cdot J_c=iT(b_I\alpha^I-b^I\beta_I)$. It is useful to rearrange all 24 non-geometric fluxes in the following matrices:
\begin{align}\label{eq:mcs}
    b_{IJ}&= \begin{bmatrix}
    	Q^{65}_1&Q^{46}_2&Q^{54}_3\\
	Q^{32}_1&Q^{34}_5&Q^{42}_6\\
	Q^{53}_4&Q^{13}_2&Q^{15}_6\\
	Q^{26}_4&Q^{61}_5&Q^{21}_3
           \end{bmatrix}, \qquad
    b^I_J= \begin{bmatrix}
    	Q^{23}_4&Q^{31}_5&Q^{12}_6\\
	Q^{26}_4&Q^{16}_2&Q^{41}_3\\
	Q^{62}_1&Q^{64}_5&Q^{24}_3\\
	Q^{35}_1&Q^{34}_2&Q^{45}_6
           \end{bmatrix}.
  \end{align}
In this manner, we can write the superpotential in the following form 
\beq
W(U,S)={\cal W}(U,S)+{\rm i}TP_3(U),
\eeq
where $P_3(U)$ is given by 
\beq\label{eq:super}
P_3(U)=\sum_{i=1}^{3}(b^0_i+{\rm i}(b_{1i}+b_{2i}+b_{3i}) U -(b^1_i+b^2_i+b^3_i) U^2 -{\rm i}b_{0i}U^3).
\eeq
%

 For the specific case of the isotropic torus $T^6$, there are some relations among the non-geometric fluxes, which in our notation read:
\begin{align}\label{eq:notationISO}
\begin{split}
b_{ij}=b_{ji} = b_\ast\,,& \quad  b^i_j=b^j_i=b^\ast\,\,\, (i \ne j)\,, \\
b_{ii} = \beta_\ast & \quad b^i_i=\beta^\ast\,, \\
b^0_i=b^0 & \quad b_{0i}=b_0\,,
\end{split}
\end{align}
with $i,j=1,2,3$. With this notation, Eq. \eqref{eq:super} recasts the form given in (\ref{eq:p3}).\\
%

 In terms of the matrices given in Eq. \eqref{eq:mcs}, the Jacobi identities ($[Z,[Z,Z]]=0$), which usually are given in the form 
  \begin{equation}
  Q\cdot H= Q^{QR}_{[M}H_{NP]}=0,
\end{equation}
can be written as
\begin{align}
\begin{bmatrix}
A\\
B
\end{bmatrix} 
\begin{bmatrix}
b_{I1}\\
b^I_1
\end{bmatrix}
=
\begin{bmatrix}
C\\
B
\end{bmatrix} 
\begin{bmatrix}
b_{I2}\\
b^I_2
\end{bmatrix}
=
\begin{bmatrix}
C\\
A
\end{bmatrix} 
\begin{bmatrix}
b_{I3}\\
b^I_3
\end{bmatrix}
= 0,
\end{align}
where
\begin{align}
A=\begin{bmatrix}
0&h_3&0&h_1&h^2&0&h^0&0\\
h^0&0&h^2&0&0&h_1&0&h_3\\
0&h^1&0&h^3&h_0&0&h_2&0\\
h_2&0&h_0&0&0&h^3&0&h^1
\end{bmatrix},
\end{align}
\begin{align}
B=\begin{bmatrix}
0&h^1&h^2&0&h_0&0&0&h_3\\
h_3&0&0&h_0&0&h^2&h^1&0\\
0&h_2&h_1&0&h^3&0&0&h^0\\
h^0&0&0&h^3&0&h_1&h_2&0
\end{bmatrix},
\end{align}
\begin{align}
C=\begin{bmatrix}
0&0&h^2&h^3&h_0&h_1&0&0\\
h_1&h_0&0&0&0&0&h^3&h^2\\
0&0&h_3&h_2&h^1&h^0&0&0\\
h^0&h^1&0&0&0&0&h_2&h_3
\end{bmatrix},
\end{align}
and similarly for $Q\cdot Q=0$. For the isotropic torus the above identities become
\begin{align}
b^0 h_0 + b^\ast h_\ast - (b_\ast + \beta_\ast) h^\ast&=0\,, \nonumber\\
b_0 h^0 + b_\ast h^\ast-(b^\ast + \beta^\ast) h_\ast &=0\,, \nonumber\\
b_\ast h_0 + b_0 h_\ast - (b^\ast + \beta^\ast) h^\ast&=0\,, \nonumber\\
b^\ast h^0+ b^0 h^\ast - (b_\ast + \beta_\ast) h_\ast&=0\,,
\label{Jacobi}
\end{align}
\begin{align}
b_\ast (b_\ast + \beta_\ast) - b^0 (b^\ast +\beta^\ast)&=0\,, \nonumber\\
b_0 (b_\ast + \beta_\ast) - b^\ast (b^\ast + \beta^\ast) &=0\,, \nonumber\\
b_0 b^0 - b_\ast b^\ast&=0\,, 
\label{QQ}
\end{align}
while the tadpole condition on the non-geometric fluxes takes the form
\beq
f_0 b^0 -  f^0 b_0  + ((2 b^\ast +\beta^\ast) f_\ast -(2 b_\ast + \beta_\ast) f^\ast)=0\,,
\eeq
for the flux conditions (\ref{eq:fluxrelations}) and the relations (\ref{solution}), meaning that seven branes are absent in our model.








